\documentclass[submission,copyright,creativecommons]{eptcs}
\providecommand{\event}{MBT 2013} 
\usepackage{breakurl}             
\usepackage{graphicx}
\usepackage{amsmath}
\usepackage{amssymb}
\usepackage{aeguill}

\DeclareMathSymbol{\B}{\mathalpha}{AMSb}{"42}
\DeclareMathSymbol{\I}{\mathalpha}{AMSb}{"49}
\DeclareMathSymbol{\N}{\mathalpha}{AMSb}{"4E}
\DeclareMathSymbol{\Pwr}{\mathalpha}{AMSb}{"50}
\DeclareMathSymbol{\Q}{\mathalpha}{AMSb}{"51}
\DeclareMathSymbol{\R}{\mathalpha}{AMSb}{"52}
\DeclareMathSymbol{\Z}{\mathalpha}{AMSb}{"5A}
\DeclareMathSymbol{\Sol}{\mathalpha}{AMSb}{"53}

\newcommand{\ist}{\mbox{{\tt true}}}

\newcommand{\dontshow}[1]{}

\newcommand{\equivdef}{\equiv_{\mbox{\footnotesize def}}}

\newcommand{\fun}{\rightarrow}

\newcommand{\currt}{\hat{t}}

\newcommand{\tx}{\mathbf{X}}
\newcommand{\tf}{\mathbf{F}}
\newcommand{\tg}{\mathbf{G}}
\newcommand{\tu}{\mathbf{U}}

\newcommand{\tw}{\mathbf{W}}

\newcounter{examplectr}
\newenvironment{example}[1] 
{
{\refstepcounter{examplectr}
\medskip
\noindent
\bf Example~\theexamplectr.\label{#1}}
}
{
\unskip\nobreak\hfil\penalty50
      \hskip2em\hbox{}\nobreak\hfil$\Box$%
      \parfillskip=0pt \finalhyphendemerits=0 \par
}

\title{Industrial-Strength Model-Based Testing - State of the Art and Current Challenges\thanks{The author's research is funded by the EU FP7 COMPASS project under grant agreement no.287829}}
\author{Jan Peleska
\institute{University of Bremen, Department of Mathematics and Computer Science, Bremen, Germany}
\institute{Verified Systems International GmbH, Bremen, Germany}
\email{jp@informatik.uni-bremen.de}
}
\def\titlerunning{Industrial-Strength Model-Based Testing}
\def\authorrunning{Jan Peleska}
\begin{document}
\maketitle

\begin{abstract}
As of today, model-based testing (MBT) is considered as leading-edge technology in industry.
We sketch the different MBT variants that -- according to our experience -- are currently applied
in practice, with special emphasis on the avionic, railway and automotive domains. The key factors
for successful industrial-scale application of MBT are described, both from a scientific
and a managerial point of view. With respect to the former view, we describe the techniques for automated
test case, test data and test procedure generation for concurrent reactive real-time systems which are
considered as the most important enablers for MBT in practice. With respect to the latter view, our 
experience with introducing MBT approaches in testing teams are sketched. Finally, the most challenging
open scientific problems whose solutions are bound to improve the acceptance and effectiveness of MBT in industry 
are discussed.
\end{abstract}

\section{Introduction}\label{sec:intro}

\subsection{Model-Based Testing}

Following the definition currently given in Wikipedia\footnote{{\tt http://en.wikipedia.org/wiki/Model-based\_testing}, (date: 2013-0211).}
\begin{quote}
``\emph{Model-based testing is application of Model based design for designing and optionally also executing artifacts to perform software testing. Models can be used to represent the desired behavior of an System Under Test (SUT), or to represent testing strategies and a test environment.''}
\end{quote}

In this definition only software testing is referenced, 
but it applies to hardware/software integration and  system testing just as well. Observe that this
definition does not require that certain aspects of testing -- such as test case identification or test procedure creation -- should be performed in an automated way: the MBT approach can also be applied manually, just as design support for testing environments, test cases and so on. This rather 
unrestricted view on MBT is consistent with the one expressed in~\cite{uml-testing-profile}, and it is
reflected by today's MBT tools ranging from graphical test case description aides to highly automated 
test case, test data and test procedure generators. 
Our concept of models also comprises computer programs, typically  represented by per-function/method control flow graphs annotated by statements and conditional expressions.

Automated MBT has received much attention in
recent years, both in academia and in industry. This interest has been
stimulated by the success of model-driven development in general, by the
improved understanding of testing and formal verification as complementary
activities, and by the availability of efficient tool support. Indeed, when
compared to conventional testing approaches, MBT has proven to increase both
quality and efficiency of test campaigns; we name~\cite{10.1109/ICST.2010.60} as one example where quantitative evaluation results have been given. 

In this paper the term model-based
testing is used in the following, most comprehensive, sense: the behaviour of
the \emph{system under test (SUT)} is specified by a model elaborated in the
same style as a model serving for development purposes. Optionally, the SUT
model can be paired with an environment model restricting the possible
interactions of the environment with the SUT. A \emph{symbolic test case
  generator} analyses the model and specifies \emph{symbolic test cases} as
logical formulas identifying model computations suitable for a certain test
purpose. Constrained by the transition relations of SUT and environment model,
a \emph{solver} computes concrete model computations which are
\emph{witnesses} of the symbolic test cases. The inputs to the SUT obtained
from these computations are used in the test execution to stimulate the
SUT. The SUT behaviour observed during the test execution is compared against
the \emph{expected} SUT behaviour specified in the original model. Both
stimulation sequences and \emph{test oracles}, i.~e., checkers of SUT
behaviour, are automatically transformed into \emph{test procedures} executing
the concrete test cases in a model-in-the-loop, software-in-the-loop, or hardware-in-the-loop
configuration.

According
to the MBT paradigm described here, the focus of test engineers is shifted
from test data elaboration and test procedure programming to modelling. The
effort invested into specifying the SUT model results in a return of
investment, because test procedures are generated automatically, and debugging
deviations of observed against expected behaviour is considerably facilitated
because the observed test executions can be ``replayed'' against the model. Moreover, V\&V processes and certification are facilitated because test cases can be automatically traced against the model which in turn reflects the complete set of system requirements.

\subsection{Objectives of this Paper}

The objective of this paper is to describe the capabilities of MBT tools which 
-- according to our experience -- are 
fit for application in today's industrial scale projects and which are essential for 
successful MBT application in practice. 
The  MBT application field   considered here is distributed embedded real-time systems in the
avionic, automotive and railway domains. 
The description refers to our tool
RT-Tester\footnote{The tool has been developed by Verified Systems International in cooperation with the author's team at the University of Bremen. It is available 
free of charge for academic research, but commercial licenses have to be obtained for industrial application. Some components (e.g., the SMT solver) will also become available as open source.} for illustrating several aspects of MBT in practice,
and the underlying methods that helped
 to meet the test-related requirements from real-world V\&V 
campaigns.  
The presentation is structured according to the MBT researchers' and tool builders' perspective: we
describe the ingredients that, according to our experience, should be present
in industrial-strength test automation tools, in order to cope with test
models of the sizes typically encountered when testing embedded real-time
systems in the automotive, avionic or railway domains.
We hope that these references to an existing tool may serve as ``benchmarking information'' which
may motivate other researchers to describe alternative methods 
and their virtues with respect to practical testing campaigns.

\subsection{Outline}

In Section~\ref{sec:rtt} a tool introduction   is given.
In Section~\ref{sec:modelling}, MBT methods and challenges related to modelling are discussed.
Section~\ref{sec:strategies} introduces a formal view on requirements, test cases and their traceability 
in relation to the test model. It also discusses various test strategies and their justification.
A case study illustrating various points of our discussion of MBT is described in Appendix~\ref{sec:turnind}.
Section~\ref{sec:conc} presents the conclusion.
We give references to alternative or competing methods and tools along the way,
as suitable for the presentation.

\section{A Reference MBT Tool}\label{sec:rtt}

RT-Tester supports all test levels from unit testing to system
integration testing and provides different functions  for manual test
procedure development, automated test case, test data and test procedure
generation, as well as management functions
for large test campaigns. The typical application scope covers (potentially
safety-critical) embedded real-time
systems involving concurrency, time constraints, discrete control decisions as
well as integer and floating point data and calculations. While the tool has
been used in industry for about 15 years and has been qualified for avionic, automotive and railway 
control systems under test according to the standards~\cite{do178b,iso26262-4,EN50128}, the results
presented here refer to more recent functionality that has been validated during the
last years in various projects from the transportation domains and are  
now made available to the public.

  \begin{figure*}[!t]
  \centering
 \includegraphics[width=7in]{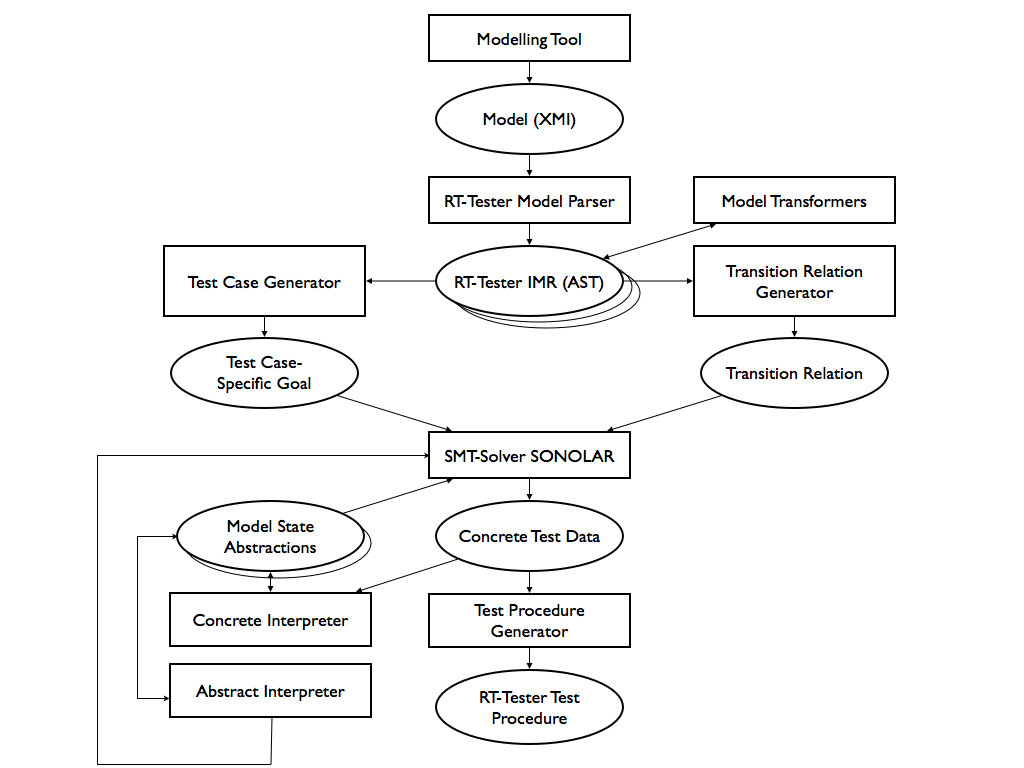}
  \caption{Components of the RT-Tester test case/test data generator.}
  \label{fig:rtttgen}
  \end{figure*}


The starting point for MBT is a concrete test model describing the expected
behaviour of the system under test (SUT) and, optionally, the behaviour of the
operational environment to be simulated in test executions by the testing
environment (TE) (see Fig.~\ref{fig:rtttgen}). 
Models developed in a
specific formalism are transformed into some textual representation supported by the
modelling tool (usually XMI format). A model parser front-end reads the model text
and creates an internal model representation (IMR) of the abstract
syntax. 

A transition relation generator creates the initial state
and the transition relation of the model as an expression in propositional logic,
referring to pre-and post-states. Model transformers create additional reduced,
abstracted or equivalent model representations which are useful to speed up
the test case and test data generation process. 

A test case   generator
creates propositional formulas representing test cases built according to a
given strategy.  A satisfiability modulo theory (SMT) solver
calculates solutions of the test case constraints in compliance with the
transition relation. This results in concrete computation fragments yielding
the time stamps and input vectors  to be used in the test procedure
implementing the test case (and possibly other test cases as well).  An
interpreter simulating the model in compliance with the transition
relation is used to investigate concrete model executions continuing the
computation fragments calculated by the SMT solver or, alternatively, creating
new computations based on environment simulation and random data
selection. An abstract interpreter supports the SMT solver in finding solutions faster by calculating the minimum number of transition steps required to reach the goal, and by restricting the ranges of inputs and other model variables for each state possibly leading to a solution.
Finally, the test procedure generator creates executable
test procedures as required by the test execution environment by mapping the
computations derived before into time-controlled commands sending input data
to the SUT and  by creating test oracles from the SUT model portion checking
SUT reactions on the fly, in dependency of the stimuli received before from
the TE.

\section{Modelling Aspects}\label{sec:modelling}

\subsection{Modelling Formalisms}

It is our expectation that the ongoing
discussions about suitable modelling formalisms for reactive systems -- from
UML via process algebras and synchronous languages
 to domain-specific languages -- will not converge to a
single preferred formalism in the near future. As a consequence it is
important to separate the test case and test data generation algorithms from
concrete modelling formalisms.

RT-Tester supports subsets of UML~\cite{UML241} and SysML~\cite{SysML12} for creating test models:
SUT structure is expressed by composite structure or block diagrams, and behaviour is specified by
means of state machines and operations (a small SysML-based case study is presented Appendix~\ref{sec:turnind}). The parser front end reads model exports from different tools in XMI
format.  Another parser reads Matlab/Simulink models. For software testing, a further front end parses transition graphs of C functions. 

The first versions of RT-Tester supported CSP~\cite{schneider00} as modelling language, but the
process-algebraic presentation style was not accepted well by practitioners. 
Support for an alternative 
 textual formalism is currently elaborated by creating a front-end for CML~\cite{woodcock2012}, 
the COMPASS
modelling language specialised on systems of systems (SoS) design, verification and validation.
In CML, the problems preventing a wider acceptance of CSP for test modelling have been 
removed.

Some formalisms are domain-specific and supported on customers' request: in~\cite{10.1109/ICST.2010.60}
automated MBT against a 
 timed variant of Moore Automata is described, which is used for modelling control logic of
 level crossing systems.

\subsection{A Sample Model}
In Appendix~\ref{sec:turnind} a case study is presented which will be used in this paper
 to illustrate
modelling techniques, test case generation and requirements tracing. The case study models
the turn indication and emergency flashing functions as present in modern vehicles. While
this study is just a small simplified example, a full test model of the turn indication 
function as realised in Daimler Mercedes cars has been published in~\cite{pel2011a} and is  
available under {\tt http://www.mbt-benchmarks.org}.

\subsection{Semantic Models}
In addition to the internal model representation which is capable of representing 
abstract syntax trees for a wide variety of formalisms, a semantic model is needed which
is rich enough to encode the different behaviours of these formalisms. 
As will be described in Section~\ref{sec:strategies}, operational model semantics
is the basis for automated test data generation, and it is also needed to specify
the conformance relation between test model and SUT, which is checked by the   
tests oracles generated from the model (see below). 

A wide variety of semantic models is available and suitable for test generation.
Different variants of labelled transition systems (LTS) are used for testing against
process algebraic models, like Hennessy's acceptance tree semantics~\cite{hennessy88}, the
failures-divergence semantics of CSP (they come in several variants~\cite{roscoe2010}) 
and Timed CSP~\cite{schneider00}, the LTS used in I/O conformance test theory~\cite{tretmansIoco96,tretmansConcur99}, or the Timed LTS used for the testing theory of
Timed I/O Automata~\cite{Springintveld2001}. As an alternative to the LTS-based approach, 
Cavalcanti and Gaudel advocate for the 
Unifying Theories of Programming~\cite{hoare1998}, 
that are used, for example, as a semantic basis for the Circus
formalism and its testing theory~\cite{cavalcant2011a}, and for the COMPASS Modelling Language CML
mentioned above.

For our research and MBT tool building foundations we have adopted Kripke Structures, mainly
because our test generation techniques are close to techniques used in (bounded) model checking,
and many fundamental results from that area are formulated in the semantic framework of 
Kripke Structures~\cite{clarke_em-etal:1999a}. Recall that a Kripke Structure is a state transition system
$K = (S, S_0, R, L)$ with state space $S$, initial states $S_0\subseteq S$, transition relation $R\subseteq S\times S$ and labelling function $L:S\fun \Pwr(AP)$ associating each state $s$ with the set $L(s)$ 
of atomic propositions $p\in AP$ which hold in this state. The behaviour of $K$ is expressed by the set of computations $\pi = s_0.s_1.s_2\ldots \in S^\omega$, that is, the infinite sequences $\pi$ 
of states fulfilling $s_0\in S_0$ and $R(s_i,s_{i+1}), i = 0,1,2, \ldots$. In contrast to LTS, Kripke Structures do not support a concept of events, these have to be modelled by propositions becoming $\ist$
when changing from one state to a successor state. For testing purposes, states $s\in S$
 are typically 
modelled by variable valuation functions $s : V \fun D$ where $V$ is a set of variable symbols $x$  mapped
by $s$ to their current value $s(x)$ in their appropriate domain (bool, int, float, \ldots) which is
a subset of $D$. The variable symbols are partitioned into $V = I \cup O \cup M$, where $I$ contains the
input variables of the SUT, $O$ its output variables, and $M$ its internal model variables which cannot
be observed during tests. Concurrency can be modelled both for the synchronous   (``true parallelism'') 
\cite{Bryant2000} and the interleaving variants of semantics~\cite[Chapter~10]{clarke_em-etal:1999a}. 
Discrete or dense time can be encoded by means of a variable $\currt$ denoting model execution time.
For dense-time models this leads to state spaces of uncountable size, but the abstractions of the
state space according to clock regions or clock zones, as known from Timed Automata~\cite{clarke_em-etal:1999a} can be encoded by means of atomic propositions and lead to finite-state abstractions.

Observe that there should   be no real controversy 
about whether LTS or Kripke Structures are more suitable for describing behavioural semantics of models:
De Nicola and Vaandrager~\cite{denicola1990} have shown how to construct property-preserving transformations of LTS into
Kripke Structures and vice versa.

\subsection{Conformance Relations}





Conformance relations specify the correctness properties of a SUT by comparing its actual
behaviour observed during test executions to the possible behaviours specified by the model.
A wide variety of conformance relations are known. For Mealy automata models, Chow used an input/output-based equivalence relation which amounted to isomorphism between minimal automata
representing specification and implementation models~\cite{chow:wmethod}.
in the domain of process algebras Hennessy and De Nicola 
introduced the relation of {\it testing equivalence} which related specification process behaviour to
SUT process behaviour~\cite{nicola84}. For Lotus, this concept was explored in 
depth by Brinksma~\cite{brinksma88}, Peleska and Siegel
 showed that it could be equally well applied for CSP and
its refinement relations~\cite{peleska1997a}, and Schneider extended these results to
Timed CSP~\cite{schneider95}. Tretmans introduced the concept of I/O conformance~\cite{tretmansIoco96}.
Vaandrager et.~al.~used bi-similarity as a testing relation between timed automata representing
specification and implementation~\cite{Springintveld2001}. All these conformance relations have in common,
that they are defined on the model semantics, that is, as relations between computations admissible for
specification and implementation, respectively.

\paragraph{Conformance in the synchronous deterministic case.}
For our Kripke structures, a simple variant of I/O conformance suffices for a surprisingly wide range
of applications: for every trace\footnote{Traces are finite prefixes of computations.} 
$s_0.s_1\ldots s_n$ identified for test purposes in the model, the associated test execution trace
$s_0'.s_1'\ldots s_n'$ should have the same length and satisfy
\[
\forall i\in \{0,\ldots,n\}: s_i|_{I\cup O\cup \{\currt\}} = s_i'|_{I\cup O\cup \{\currt\}}
\]
that is, the observable input and output values, as well as the time stamps should be identical.

This very simple notion of conformance is justified for the following scenarios of reactive systems testing:
(1) The SUT is non-blocking on its input interfaces, (2) the  most recent value passed along output
interfaces can always be queried in the testing environment, (3) each concurrent component is deterministic, and (4) the synchronous concurrency semantics applies. At first glance, these conditions may seem rather restrictive, but there is a wide variety of practical test applications where they apply:
many SUT never refuse inputs, since they communicate via shared variables, dual-ported ram, or non-blocking
state-based protocols\footnote{In the avionic domain, for example, the sampling mode of the AFDX protocol \cite{ARINC664P7-1}
allows to transmit messages in non-blocking mode, so that the receiver always reads the most recent data
value.}. Typical hardware-in-the-loop testing environments always keep the current output values of the SUT 
in memory for evaluation purposes, so that even message-based interfaces can be accessed as shared variables
in memory (additionally, test events may be generated when an output message of the SUT actually arrives in the test environment (TE). For safety-critical applications the control decisions of each sequential sub-component of the SUT must be deterministic, so that the concept of {\it may tests}~\cite{hennessy88}, where a test trace may or may not be refused by the SUT does not apply. As a consequence, the complexity and elegance 
of testing theories handling non-deterministic internal choice
and associated refusal sets and unpredictable outputs of the SUT are not applicable for these types of
systems. Finally, synchronous systems are widely used for local control applications, such as, 
for example, PLC controllers or computers adhering to the cyclic processing paradigm.

In RT-Tester this conformance relation is practically applied, for example, when testing
software generated from SCADE models~\cite{scade}: the SCADE tool and its modelling language adhere
 to the synchronous paradigm. The software operates in processing cycles. Each cycle starts with
 reading input data from global variables shared with the environment; this is followed by internal processing steps, and the output variables are updated at the end of the cycle. Time $\currt$ is a discrete abstraction
 corresponding to a counter of processing cycles.

\paragraph{Conformance in presence of non-determinism.}
For distributed control systems the synchronous paradigm obviously no longer applies,
and though single sequential SUT components will usually still act in a deterministic way, 
their outputs will interleave non-deterministically with those of others executing in a concurrent way.
Moreover, certain SUT outputs may change non-deterministically over a period of time, because 
the exact behavioural specification is unavailable. These aspects are supported in RT-Tester
 in the following ways.
 \begin{itemize}
 \item All SUT output interfaces $y$ are associated with (1) an acceptable deviation $\varepsilon_y$
 from the accepted value (so any observed value $s'(y)$ deviating from the expected value $s(y)$ by 
 $|s'(y) - s(y)| \leq \varepsilon$ is acceptable), (2) an admissible latency $\delta_y^0$ (so  
 any observed value $s'(y)$  for the expected value $s(y)$ 
 is not timed out as long as $s'(\currt) - s(\currt) \leq \delta_y^0$, 
 and (3) an acceptable time $\delta_y^1$
 for early changes of $y$ (so $s(\currt) - s'(\currt) \leq \delta_y^1$ is still acceptable).

\item A time-bounded non-deterministic assignment statement {\tt y = UNDEF(t,c)} stating that 
$y$'s valuation is arbitrary for a duration of {\tt t} time units, after which it should assume
value {\tt c} (with an admissible deviation and an admissible delay).

\item A model transformation turning the SUT model into a test oracle: it
\begin{itemize}
\item extends the variable space by one additional output variable $y'$ per SUT output $y\in O$,
\item adds one concurrent  checker component ${\cal O}_y$
 per SUT output signal, operating on $y$ and $y'$,
\item adds one concurrent component ${\cal P}$ processing the timed input output trace as observed during
the test execution, with observed SUT outputs written to $y'$ (instead of $y$),
\item transforms each concurrent SUT component $C_i$ into ${\cal C}_i$.
\end{itemize}
This is described in more detail in the next paragraphs. 
\end{itemize}

\begin{figure}
  \centering
 \includegraphics[width=5in]{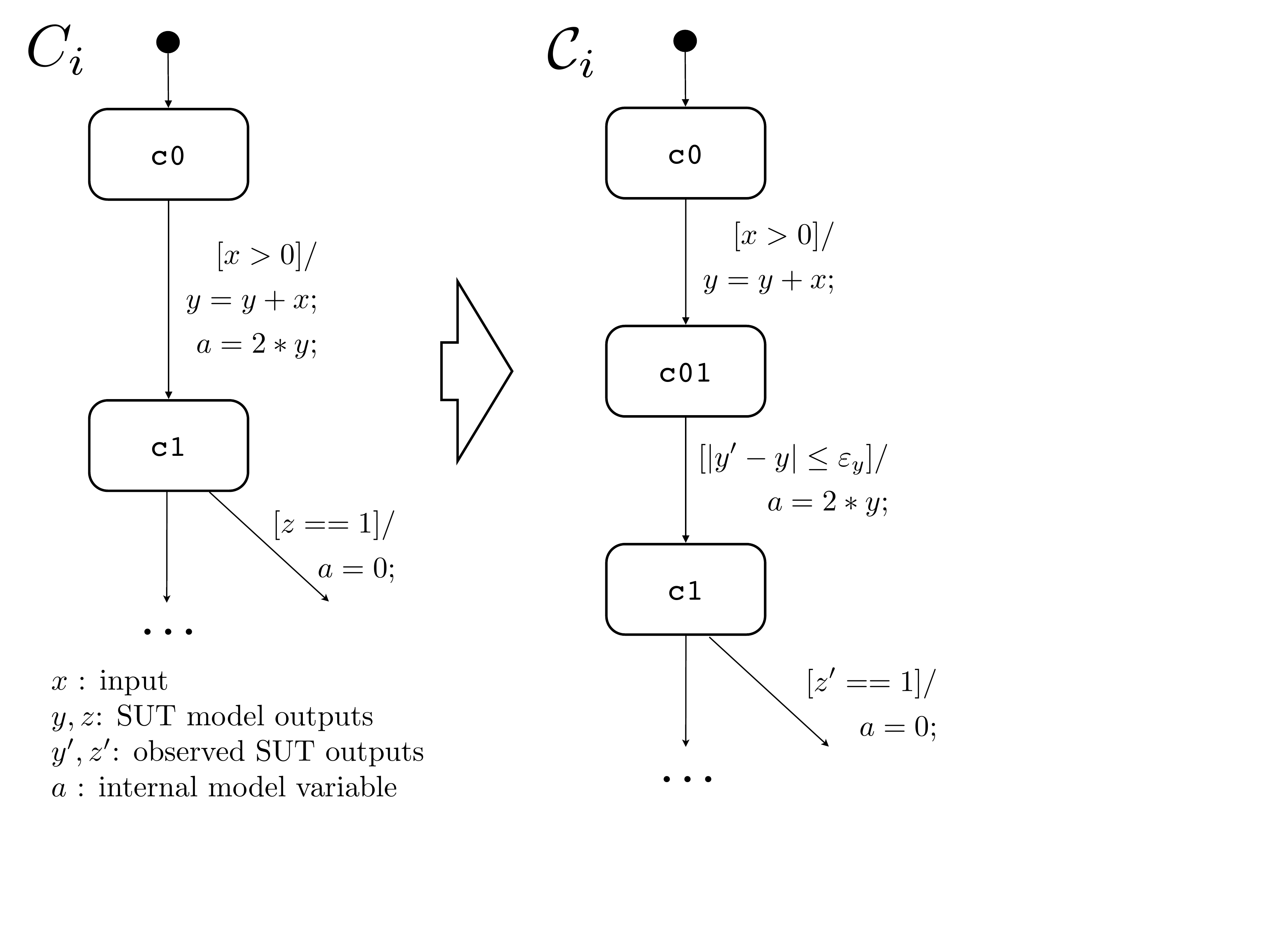}
 \vspace*{-10mm}
  \caption{Example of original SUT component $C_i$ and transformed component ${\cal C}_i$.}
  \label{fig:csut}
  \end{figure}

The transformed SUT components ${\cal C}_i$ operate as sketched in the example shown 
in Fig.~\ref{fig:csut}. Every write of $C_i$ to some output $y$ is performed in
${\cal C}_i$ as well, ${\cal C}_i$ however, waits for the corresponding output value $y'$ 
observed during test execution to change until it fits to the expected value of $y$ (guard
condition $|y'-y| \leq\varepsilon$). This helps to adjust to small admissible delays of in the
expected change of $y'$ observed in the test: the causal relation ``$a$ is written after $y$ has been
changed is preserved in this way.
If $C_i$ uses another output $z$ (written, for example,
by a concurrent component $C_j$) in a guard condition, it is replaced by variable $z'$ containing
the observed output during test execution. This helps to check for correctness of
 relative time distances like 
``output $w$ is written 10ms after $z$ has been changed'', if the actual output on $z'$ is delayed by an admissible amount of time.

\begin{figure}
  \centering
 \includegraphics[width=5in]{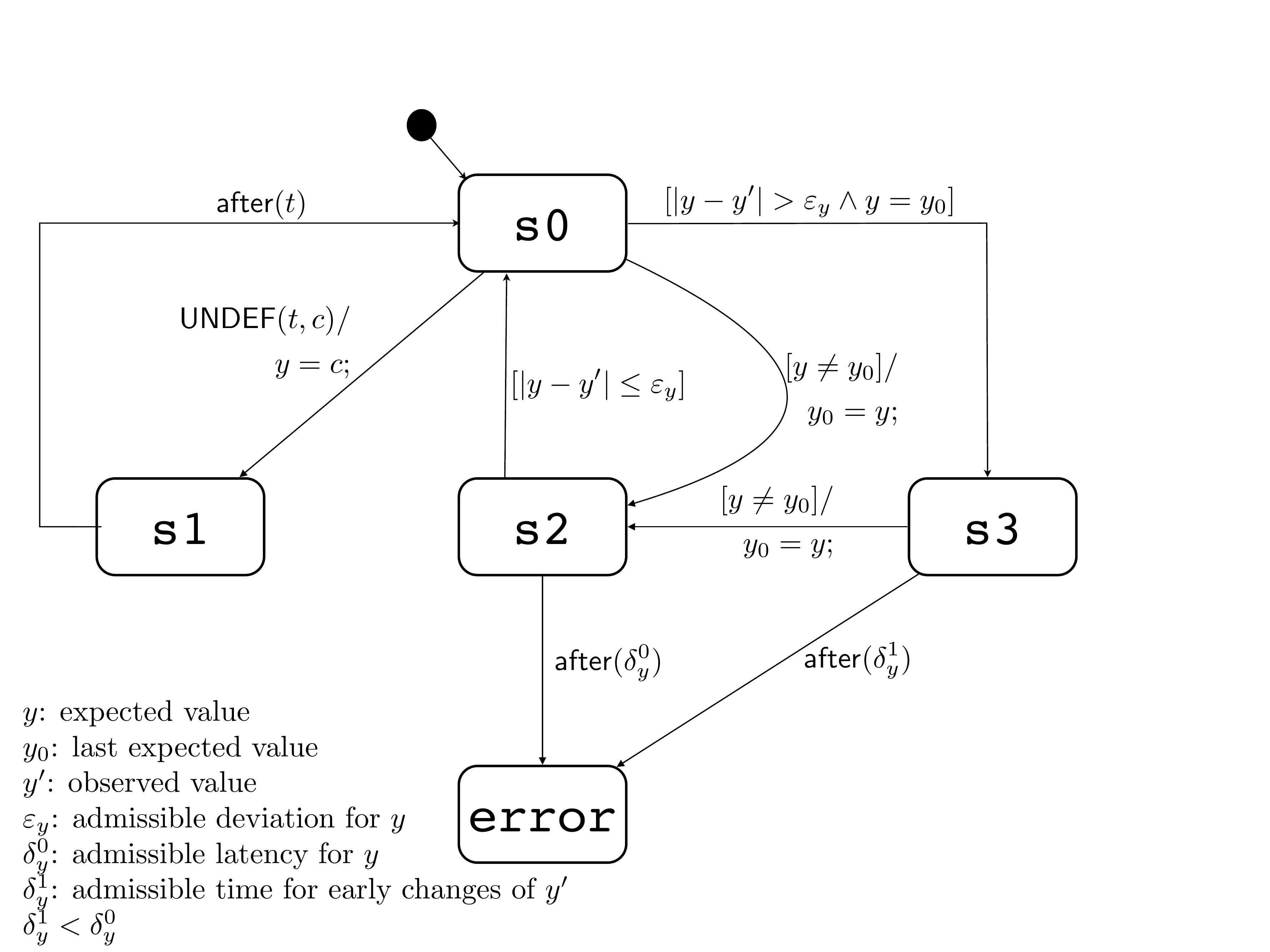}
  \caption{Test oracle component observing one SUT output interface $y$.}
  \label{fig:testora}
  \end{figure}

The concurrent test oracles ${\cal O}_y$ operate as shown in Fig.~\ref{fig:testora}: 
If some component ${\cal C}_i$ writes to   an expected output $y$, the oracle traverses into
control state {\tt s2}. If the corresponding observed output $y'$ is also adjusted in ${\cal P}$, such
that $|y'-y| \leq \varepsilon_y$ holds before $\delta_y^0$ time units have elapsed, the change to $y'$ 
is accepted and the oracle transits to {\tt s0}. Otherwise the oracle transits into the  error state.
If
the observed value changes unexpectedly above threshold $\varepsilon_y$, the oracle changes into 
location {\tt s3}. If the expected value $y$ also changes shortly afterwards, this means that 
the SUT was just some admissible time earlier than expected according to the model, and the change
is monitored via state {\tt s2} as before. If $y$, however, does not change for at least $\delta_y^1$ 
time units, we have uncovered an illegal output change of the SUT and transit into the error state.

A test execution (that is, an input/output trace) 
performed with the SUT conforms to the model if and only if the
transformed model accepts the test execution processed by ${\cal P}$ in the sense that
none of the oracles transits into an error state.
RT-Tester uses this conformance relation   for hardware-in-the-loop system
testing, as, for example,  in the tests of the automotive controller network supporting the turn indication function
in Daimler Mercedes vehicles~\cite{pel2011a}.







\subsection{Test-Modelling Related Challenges}







With suitable test models available, test efficiency and test quality are improved in a considerable
way. The elaboration of a model, however, can prove to be a major hurdle for the success of MBT in practice.
\begin{enumerate}
\item If complex models have to be completed before testing can start, this induces  an unacceptable delay 
for the proper test executions.
\item For complex SUT, like systems of systems, test models need to abstract from a large amount of detail,
because otherwise the resulting test model would become unmanageable.
\item The required skills for test engineers writing test models are significantly  higher
than for test engineers writing sequential test procedures.
\end{enumerate}

We expect that   problem 1 will be solved in the future by 
incremental model development, where test suites with increasing coverage and error detection capabilities
 can be run between model increments. The current methods based on sequential state machines as described by \cite{Vaandrager12} may be extended to partially automated approaches where test model designers provide -- apart from interface descriptions -- initial architectural frames and suggestions for internal state variables, and automated machine learning takes these information into account. Furthermore, the explicit state machine construction may be complemented by incremental elaboration of transition relations: as pointed out by \cite{pelnfm2011} for the purpose of test data generation, concurrent real-time models with complex state space are often better expressed by means of their transition relation than by explicit concurrent state machines.
  Promising attempts to construct test models in an incremental way from actual observations obtained during SUT simulations or experiments with the actual SUT indicate that test model development can profit from  ``re-engineering'' SUT properties or model fragments from observations~\cite{rogin:338}.  
  
  The problem of model complexity can be overcome by introducing contracts for the constituent systems of
  a large system of systems. This type of abstractions is investigated, for example, in the COMPASS project\footnote{{\tt http://www.compass-research.eu}}.
  
  With respect to the third problem it is necessary to point out in management circles that competent 
  testing requires the same skills as competent software development. So if modelling skills are required 
  for model-driven software and system development, these  skills are required for test engineers as well.

\section{Requirements, Test Cases and Trustworthy Test Strategies}\label{sec:strategies}

\subsection{Requirements}
If a test model has been elaborated in an adequate way, it will reflect the requirements
to be tested. At first glance, however, it may not be obvious to identify the model
portions contributing to a given requirement. Formally speaking, a requirement is reflected
by certain computations $\pi = s_0.s_1.s_2\ldots$ of the model. Computations can be identified,
for example, by some variant of temporal logic, 
and we use Linear Temporal Logic (LTL)~\cite[Chapter~3]{clarke_em-etal:1999a}
for this purpose\footnote{Recall that LTL uses 4 path operators: $\tg\phi$ (globally $\phi$) states
that $\phi$ holds in every state of the computation. $\tf\phi$ (finally $\phi$) states that $\phi$ holds
in some computation state. $\tx \phi$ states that $\phi$ holds in the next state following
the computation state under consideration. $\phi\ \tu\ \psi$ states that finally 
$\psi$ will hold in a computation state and $\phi$ fill hold in all previous states (if any).}.

Consider, for example,  requirement {\sf REQ-001} (Flashing requires sufficient voltage)
from the sample application specified in Appendix~\ref{sec:turnind}, Table~\ref{tab:req}.
It can be readily expressed in LTL as  
\begin{equation}\label{eq:req001}
\tg (\text{Voltage} \leq 80 \Rightarrow 
\tx (\neg(\text{FlashLeft}\vee\text{FlashRight})\ \tu\ \text{Voltage} > 80))
\end{equation}

This is a black-box specification: it only refers to input and output interfaces of the SUT
and is valid without any model. With a model at hand, however, the specification can be slightly
simplified, because the relevant SUT reactions have been captured by state machine OUTPUT\_CTRL  
(see Fig.~\ref{fig:OUTPUT_CTRL})\footnote{Control states are encoded as Boolean variables
in the model state space, $\text{Idle} = \ist$ means that state machine OUTPUT\_CTRL is in 
control state $\text{Idle}$.}.
\[
\tg (\text{Voltage} \leq 80 \Rightarrow 
\tx (\text{Idle}\ \tu\ \text{Voltage} > 80))
\]
In control state Idle the indication lights are never activated. Now the computations contributing
to {\sf REQ-001} are exactly the ones finally fulfilling the premise $\text{Voltage} \leq 80$, where
the effect of the requirement may become visible, that is,
\[
\tf (\text{Voltage} \leq 80)
\]
It is unnecessary to specify the effects of the requirement in this formula, because we are only 
considering valid model computations, and the  effect is encoded in the model.

Observe that the application of LTL to characterise model computations associated with a requirement
 differs from its 
utilisation for black-box specification as in formula (\ref{eq:req001}), where the behaviour required 
along those computations has to be specified in the formula, and only interface variables of the
system may be referenced. It also differs from the application
of temporal logics  in property checking, where either all (a required property) 
or no computations (a requirements violation) of the model should
fulfil the formula.

Referring to internal model elements frequently simplifies the formulas for characterising computations.
Requirement {\sf REQ-002} (Flashing with 340ms/320ms on-off periods), for example, is witnessed by all
computations satisfying (see Fig.\ref{fig:FLASHING})
\begin{equation}\label{eq:req002b}
\tf(\text{OFF} \wedge \tx \text{ON})
\end{equation}

\subsection{Requirements Tracing to the Model}

The SysML modelling formalism~\cite{SysML12} provides syntactic means to identify requirements
in the model.   In Fig.\ref{fig:FLASHING}, for example, the transitions $\text{ON} \fun \text{OFF}$ and
$\text{OFF} \fun \text{ON}$ realise the flashing period specified by {\sf REQ-002}. This is documented
by means of the {\sf\guillemotleft satisfy\guillemotright} relation drawn from the transitions to the requirement. The interpretation
of this relation is that every model computation finally covering one of the two transitions or both contributes to the requirement. Since   computations cover $\text{OFF} \fun \text{ON}$
if and only if they fulfil $\tf(\text{OFF} \wedge \tx \text{ON})$, the {\sf\guillemotleft satisfy\guillemotright} 
relation from $\text{ON} \fun \text{OFF}$ to {\sf REQ-002} is redundant. Other examples for such simple
relationships between model elements an requirements are shown in the state machine 
depicted in Fig~\ref{fig:EMER_ON}. Formally speaking, these simple relationships are of the type
\begin{equation}\label{eq:stateform}
\tf \langle\text{State Formula}\rangle
\end{equation}
where the state formula expresses the condition that a model   element related to the 
requirement is covered: for {\sf REQ-002}, the formula  (\ref{eq:req002b}) can be expressed in the 
form (\ref{eq:stateform}) as
\[
\tf (\text{OFF} \wedge (\currt - t_{\text{OFF}}) \geq 320
\]
Here $t_{\text{OFF}}$ denotes the timer variable that stores the current time whenever control state
$\text{OFF}$ is entered and $\currt$ is the current model execution time, 
so $(\currt - t_{\text{OFF}})$ expresses   the fact that the relative time event $\text{after}(320ms)$
has occurred. In this case the transition $\text{OFF} \fun \text{ON}$ must be taken, since UML/SysML
state machine priority assigns higher priority to lower-level transitions: even if transitions
$\text{FLASHING} \fun \text{FLASHING}$ or $\text{FLASHING} \fun \text{Idle}$ of the state machine
in Fig.~\ref{fig:OUTPUT_CTRL} are enabled, transition  $\text{OFF} \fun \text{ON}$ has higher priority
because it resides in the sub-maschine of $\text{FLASHING}$.

Evaluations of system requirements in the automotive domain (in cooperation with Daimler) have shown 
that approximately 80\% of requirements are reflected by model computations satisfying
\[
\tf\left(\bigvee_{i=0}^h\phi_i \right) 
\]
where the $\phi_i$ are state formulas, each one expressing coverage of a single model element.

About 20\% of system requirements require more complex witnesses, whose LTL specification
involve nested path operators and  state formulas
referring to model elements, variable valuations and time. For these situations, we use constraints
containing the more complex LTL formulas, and the constraints are linked to their associated requirements
by means of the {\sf\guillemotleft satisfy\guillemotright} relation. Table~\ref{tab:tracing} lists the requirements of the
case study captured in Table~\ref{tab:req}, and associates the constraints characterising the witness 
traces for each requirement.

\subsection{Test Cases}

Since tests must terminate after a finite number of steps, they consist of traces $\iota = s_0\ldots s_k$
probing prefixes of relevant model computations $\pi = s_0\ldots s_k.s_{k+1}\ldots$.
If $\pi$ is a witness for some requirement {\sf R} characterised by LTL formula $\phi$, a suitable
test case $\iota$ has to be constructed in a way that at least does not violate $\phi$ while 
transiting through states $s_0\ldots s_k$, even though $\phi$ will be violated by many possible 
extensions of  $\iota$. This problem is well-understood from the field of bounded model checking (BMC),
and Biere et.~al.~\cite{BiereEA:BMC,biere2006} introduced a step semantics for evaluating
LTL formulas on finite model traces. To this end,
expression $\langle \varphi \rangle_i^{k-i}$ states that formula $\varphi$ 
holds in state $s_i$ of  a trace of 
  length $k+1$. For the operators of  LTL, their semantics can then be specified 
  inductively by\footnote{The semantics presented in~\cite{biere2006} has been simplified for our purposes.
  In~\cite{biere2006}, the authors consider possible cycles in the transition graph which are reachable
  within a bounded number of steps from $s_0$. This is used to prove the existence of witnesses for 
  formulas whose validity can only be proven on infinite paths. For testing purposes, we are only dealing with finite traces anyway; this leads to the slightly simplified bounded step semantics presented here.}
        \begin{itemize}
                \item $\langle \mathbf{G}~\varphi \rangle_0^k
        = \bigwedge_{i=0}^k \langle \varphi \rangle_i^{k-i}$ \ \ \ ($\tg \phi$ is not violated on 
        $\iota = s_0\ldots s_k$)
        
                \item $\langle \mathbf{X}~\varphi \rangle_i^{k-i}
        = \langle \varphi \rangle_{i+1}^{k-i - 1}$
        
                \item $\langle \varphi~\mathbf{U}~\psi \rangle_i^{k-i}
        = \langle \psi \rangle_i^{k-i} \vee (\langle\varphi\rangle_i^{k-i} \wedge \langle \varphi~\mathbf{U}~\psi \rangle_{i+1}^{k-i-1})$, \ \ \ \ $\langle\tf \psi\rangle_i^{k-i} = \langle \ist~\mathbf{U}~\psi \rangle_i^{k-i}$
        \end{itemize}
        
Using this bounded step semantics, each LTL formula can be transformed into formulas         
of the type
\begin{equation}\label{eq:bmc}
tc \equiv J(s_0) \wedge \bigwedge_{i=0}^n \Phi(s_i,s_{i+1}) \wedge
G(s_0,\ldots,s_{n+1})
\end{equation}
which we call {\it symbolic test cases}\footnote{In the context of BMC, these formulas
are called {\it bounded model checking instances}.} and which can be handled by the SMT solver.
Conjunct $J(s_0)$ characterises 
the current model state $s_0$ 
from where the next test objective represented by some LTL formula $\phi$ 
should be covered. This formula has to be translated into a predicate $G(s_0,\ldots,s_{n+1})$, using
the semantic rules listed above.
Predicate $\Phi$ is the transition relation of the model, and
conjunct $\bigwedge_{i=0}^n \Phi(s_i,s_{i+1})$  ensures that the solution of  $G(s_0,\ldots,s_{n+1})$ results 
in a valid trace of the model, starting from $s_0$. 

\begin{example}{ex:ltl}
Consider LTL formula 
\[
  \phi \equiv (x = 0) \tu (y > 0 \wedge \tx (\tg z = 1))
\]         
and suppose we are looking for a witness trace $\iota = s_0\ldots s_n \ldots$ with a length  of at least
$n+1$ or longer. 
Then the SMT solver is activated with the following BMC instances to solve.

\noindent
In step~0, try solving
\[
  bmc_0 \equiv \left(\bigwedge_{i=0}^n \Phi(s_i,s_{i+1})\right) \wedge s_0(y) > 0 \wedge  
  \left(\bigwedge_{i=1}^{n+1} s_i(z) = 1 \right) 
\]   
If this succeeds we are done:  the solution of $bmc_0$ is a legal trace $\iota$ of the model, since
$\Phi(s_i,s_{i+1})$ holds for each pair of consecutive states in $\iota$. 
Formula $\phi$ holds on $\iota$ because $y > 0$ is true in $s_0$ and
$z = 1$ holds for states $s_1\ldots s_{n+1}$, so the right-hand side operand of $\tu$ is fulfilled in the initial state of this trace.

\noindent
Otherwise we try to get a witness for the following formula in step~1.
\[
    bmc_1 \equiv \left(\bigwedge_{i=0}^n \Phi(s_i,s_{i+1})\right) \wedge s_0(x) = 0 \wedge  
    s_1(y) > 0 \wedge
  \left(\bigwedge_{i=2}^{n+1} s_i(z) = 1 \right)
\]  
          
\noindent
If no solution exists we continue with step~2.
\[
      bmc_2 \equiv \left(\bigwedge_{i=0}^n \Phi(s_i,s_{i+1})\right) \wedge s_0(x) = 0 \wedge s_1(x)= 0 \wedge  
    s_2(y) > 0 \wedge
  \left(\bigwedge_{i=3}^{n+1} s_i(z) = 1 \right)
\]
and so on, until a solution is found or no solution of length $n+1$ is feasible.
\end{example}                 

While   LTL formulas are well-suited to specify computations fulfilling a wide variety of constraints,
it has to be noted that it is also capable of defining properties of computations that will never
be tested in practice, because they can only be verified on infinite computations and not on 
finite trace prefixes thereof  (e.g., fairness properties). 
It is therefore desirable to identify a subset of LTL formulas 
that are tailored to the testers' needs for specifying finite traces with certain properties. 
This subset is called {\it SafetyLTL} and   has been introduced in~\cite{Sis94}. It 
is suitable for defining safety 
properties of computations, that is, properties that can always be falsified on a finite 
computation prefix. The SafetyLTL subset of LTL can be syntactically characterised as follows.
\begin{itemize}
                \item Negation is 
                only allowed before atomic propositions (so-called {\it negation normal form}).
                \item Disjunction $\vee$ and conjunction $\wedge$ are always allowed.  
                \item Next operators $\tx$, globally operators $\tg$ and 
                 weakly-until operators $\tw$   are allowed\footnote{Recall that the weakly-until operator is defined as $\phi~\tw~\psi \equivdef (\phi~\tu~\psi) \vee \tg \phi$, and that the 
                 until operator can be expressed by $\phi~\tu~\psi \equiv (\phi~\tw~\psi) \wedge \tf \psi$.}.
                \item Semantically equivalent formulas also belong to SafetyLTL.
        \end{itemize}

Concrete test data is created by solving constraints of the type displayed in Equation~(\ref{eq:bmc})
using the integrated SMT solver SONOLAR~\cite{pelnfm2011}. 
Finally the test procedure generator takes the solutions calculated by the SMT solver and turns them into stimulation sequences, that is, timed input traces to the SUT. Moreover, the test procedure generator
creates test oracles from the model components describing the SUT behaviour.

In requirements-driven testing, $G(s_0,\ldots,s_{n+1})$
specifies traces that are {\it witnesses} of a certain requirement {\sf R}. 
Indeed, Formula~(\ref{eq:bmc}) specifies
an {\it equivalence class} of traces that are suitable for testing {\sf R}. In model-driven testing,
$G(s_0,\ldots,s_{n+1})$ specifies traces that are suitable for covering certain portions (control states, 
transitions, interfaces, \ldots) of the model. In the paragraphs below it will be explained how 
requirements-driven and model-driven testing are related to each other.

%

\subsection{Model Coverage Test Cases}

Since adequate test models express all SUT requirements to be tested, it is reasonable
to specify and perform test cases achieving model coverage. As we have seen above, a behavioural 
model element (state machine control state, transition, operation, \ldots)
is covered by a trace $\iota = s_0\ldots s_k$,
 if the element's behaviour is exercised during some transition
$s_i \fun s_{i+1}$. For a control state $c$ this means that $s_{i+1}(c) = \ist$, and, consequently, 
the state's entry action (if any) is executed. For a transition this means that its firing condition
becomes true in some $s_i$. Operations $f$ are covered when they are associated with actions
of covered states or transitions executing $f$.

There exists a wide variety of model coverage strategies, many of them are discussed in~\cite{weissleder:diss}. The standards for safety-critical systems development and V\&V 
have only recently started to consider the model-driven development and V\&V paradigm. 
It seems that the avionic standard RTCA DO-178C~\cite{DO178C} is currently the most advanced 
with respect to model-based systems engineering. It   requires  to achieve
operation coverage, transition coverage, decision coverage, and equivalence class and boundary value 
coverage, when verifying design models~\cite[Table MB.6-1]{DO331}. Neither the standard, nor  \cite{weissleder:diss},   however,
elaborate on coverage of timing conditions (e.g., clock zones in Time Automata) or 
the coverage of execution state vectors of concurrent model components.

In RT-Tester, the following model coverage criteria are currently implemented: (1) basic control state 
coverage, (2) transition coverage, MC/DC coverage, (3) hierarchic transition coverage\footnote{This applies
to higher-level transitions of hierarchic state machines: they are exercised several times 
with as many subordinate control states as possible.} with or without MC/DC
coverage, (4) equivalence class and boundary value coverage, (5) basic control state pairs coverage, (6) interface coverage and (7) block coverage.

Basic control state pairs coverage exercises all feasible control state combinations of concurrent 
state machines in writer-reader relationship.
The equivalence class coverage technique in combination with basic control state pairs coverage
also produces a (not necessarily complete) coverage of clock zones.

Each of these coverage criteria can be specified by means of LTL formulas or, equivalently, BMC instances.

\begin{example}{ex:cspairs}
For   state machine FLASH\_CTRL (Fig.~\ref{fig:FLASH_CTRL}), the hierarchic transition coverage is
achieved by test cases 
\begin{eqnarray*}
tc_1 & \equiv & \tf (\text{EMER\_OFF} \wedge \text{EmerFlash})
\\
tc_2 & \equiv & \tf (\text{EMER\_ACTIVE} \wedge \text{TurnIndLvr} \neq 0 \wedge {}
\\ & &
((\text{TurnIndLvr} = 1) \neq \text{Left1} \vee (\text{TurnIndLvr} = 2) \neq \text{Right1}))
\\
tc_3 & \equiv & \tf (\text{EMER\_ACTIVE} \wedge (\text{Left1}\vee \text{Right1})
\wedge \text{TurnIndLvr} = 0) 
\\
tc_4 & \equiv & \tf (\text{TURN\_IND\_OVERRIDE} \wedge  \text{TurnIndLvr} = 0)
\\
tc_5 & \equiv & \tf (\neg\text{EmerFlash} \wedge \text{EMER\_ACTIVE} \wedge {}
\\ & & 
((\text{TurnIndLvr} \neq 0 \wedge \text{TurnIndLvr} = \text{Left1}\vee \text{TurnIndLvr} = \text{Right1})
\vee {}
\\ & &
(\text{TurnIndLvr} = 0 \wedge \neg (\text{Left1}\vee \text{Right1})))
\\
tc_6 & \equiv  & 
 \tf (\neg\text{EmerFlash} \wedge \text{TURN\_IND\_OVERRIDE} \wedge  \text{TurnIndLvr} \neq 0)
\end{eqnarray*}
\end{example}


\subsection{Automated Compilation of Traceability Data}

Having identified the test cases suitable for model coverage, these can be related to
requirements in an automated way.
\begin{itemize}
\item If   requirement {\sf R} is linked to model elements by   {\sf\guillemotleft satisfy\guillemotright} relationships,
then the test cases covering  these elements are automatically related to {\sf R}.

\item If requirement {\sf R} is characterised by a   LTL formula $\phi$ not directly
related to model elements, we proceed as follows.
\begin{itemize}
\item Transform $\phi$ into disjunctive normal form $\phi \equiv \bigvee_{i=0}^m \phi_i$ and 
associate test cases for each $\phi_i$ separately.

\item Each test case $tc \equiv \psi$ derived from the model is related to {\sf R},   if
$\psi \Rightarrow \phi_i$ holds.

\item If test case $tc \equiv \psi$ is neither stronger nor weaker 
 than the requirement in the sense that $\psi \wedge \phi_i$ has a solution, add a new test case
 $tc' \equiv \psi \wedge \phi_i$ and relate $tc'$ to {\sf R}. 
 
 \item If at least one of two test cases 
 $tc_1 \equiv \tf \psi_1$ and $tc_2 \equiv \tf \psi_2$ implies the requirement
 and $ tc' \equiv \tf(\psi_1 \wedge \psi_2)$ has a solution, add $tc'$ to the test case database and 
 trace it to {\sf R}.
\end{itemize}
\end{itemize}

\begin{example}{ex:tcreqtrace}
Consider requirement {\sf REQ-002} (Flashing with 340ms/320ms on-off periods) of the
example from Table~\ref{tab:req}. It is characterised by covering transitions $\text{ON} \fun \text{OFF}$ 
and  $\text{OFF} \fun \text{ON}$ (see Table~\ref{tab:tracing}). By tracing these transitions back
to model coverage test cases, the following cases can 
 be identified, and these trace back to  {\sf REQ-002}.
\begin{eqnarray*}
tc_7 & \equiv & \tf (\text{OFF} \wedge (\currt - t_{\text{OFF}}) \geq 320)
\\
tc_8 & \equiv &  \tf (\text{OFF} \wedge (\currt - t_{\text{OFF}}) \geq 320 \wedge \text{TurnIndLvr} = 1)
\\ 
tc_9 & \equiv &  \tf (\text{OFF} \wedge (\currt - t_{\text{OFF}}) \geq 320 \wedge \text{TurnIndLvr} = 2)
\\ 
tc_{10} & \equiv &  \tf (\text{OFF} \wedge (\currt - t_{\text{OFF}}) \geq 320 \wedge \text{EMER\_ACTIVE})
\\ 
tc_{11} & \equiv &  \tf (\text{OFF} \wedge (\currt - t_{\text{OFF}}) \geq 320 \wedge \text{TURN\_IND\_OVERRIDE})
\end{eqnarray*}
\end{example}
The test cases listed here are only a subset of the complete list that traces back to {\sf REQ-002}.
Test cases $tc_8, tc_9$ result from combining interface coverage on SUT input TurnIndLvr with
coverage of the $\text{OFF} \fun \text{ON}$. Cases $tc_{10},tc_{11}$ result from combining 
basic control state pairs coverage with the transition coverage. Test case $tc_7$ is redundant if any of the 
others is performed. It is quite obvious that the test case generation technique defined above 
runs into combinatorial explosion problems. Even for the small sample system discussed here, the list
of test cases from Example~\ref{ex:tcreqtrace} could be extended by 
\begin{eqnarray*}
tc_{12} & \equiv &  \tf (\text{OFF} \wedge (\currt - t_{\text{OFF}}) \geq 320 \wedge \text{EMER\_ACTIVE}
\wedge \text{TurnIndLvr} = 0)
\\ 
tc_{13} & \equiv &  \tf (\text{OFF} \wedge (\currt - t_{\text{OFF}}) \geq 320 \wedge \text{EMER\_ACTIVE}
\wedge \text{TurnIndLvr} = 1)
\\
tc_{14} & \equiv &  \tf (\text{OFF} \wedge (\currt - t_{\text{OFF}}) \geq 320 \wedge \text{EMER\_ACTIVE}
\wedge \text{TurnIndLvr} = 2)
\\
& & \ldots
\end{eqnarray*}

\subsection{Test Case Selection According to Criticality}

It is quite obvious that the number of test cases related to a requirement can become
quite vast, and that some of the test cases investigate more specific situations than others.
This problem is closely related to the problem of exhaustive testing which will be discussed below.
Since an exhaustive execution of all test case combinations related to a requirement 
will be impossible for fair-sized systems, a justified reduction of the potentially applicable 
test cases to a smaller collection is required. In the case of safety-critical systems development,
such a justification should conform to the standards  applicable for V\&V of these systems.

In the case of avionic systems, the RTCA DO-178C standard~\cite{DO178C} requires structural tests
with respect to data and control coupling and full requirements 
coverage   through testing, but does not specify
when a requirement has been verified with a sufficient number of test cases. Instead, the standard
gives test end criteria by setting code coverage goals, the coverage to be achieved depending on
the SUT's criticality~\cite[MB.C-7]{DO331}: for assurance 
level 1 systems (highest criticality), MC/DC coverage has to be achieved, for level 2 decision coverage, and for level 3 statement coverage. For levels 4 and 5, only high-level requirements have to be
covered without setting any code coverage goals, and for assurance level 5 the requirement to test data 
and control coupling is dropped. 

As a consequence, the model-based test case coverage can be tuned according to the code coverage achieved,
whenever the source code is available and the assurance level is in 1 --- 3: start with basic control state
coverage cases related to the requirement, increase coverage by adding hierarchic and MC/DC coverage test 
cases until the required code coverage is achieved. Add interface and basic control state pairs coverage cases until the data and control coupling coverage has been achieved as well.
For levels 4 or 5, no discussion is necessary, since here any ``reasonable'' test case assignment to each high-level requirement
is acceptable, due to the low criticality of the SUT. 

When MBT is applied on system level, however, it will generally be infeasible to measure code coverage
achieved during system tests. For systems of systems, in particular, system-level tests will never
cover any significant amount of code coverage, and the coverage values achieved will not be obtainable
in most cases, both for technical and for security reasons. Here we suggest to proceed as follows.

\begin{itemize}
\item For assurance level 3, exercise 
\begin{itemize}
\item interface tests -- this ensures verification of data and control coupling,
\item basic control state coverage test cases,
\item refine these test cases $tc \equiv \psi$ only if requirements have stricter characterisations
$\phi_i$; in this case add $tc' \equiv \psi\wedge \phi_i$. 
\end{itemize}

\item For assurance level 2, follow the same pattern, but use transition coverage test cases.

\item For assurance level 1, exercise
\begin{itemize}
\item interface tests,
\item basic control state pairs coverage test cases to refine the data and control coupling tests (recall
that these test cases stem from writer-reader analyses),
\item MC/DC coverage test cases in combination with hierarchic transition coverage,
\item first-level refinements of test cases related to requirements    as 
illustrated in Example~\ref{ex:tcreqtrace},
\item second level refinements (as in test cases $tc_{12}, tc_{13}, tc_{14}$ above), if 
the additional conjuncts have direct impact on the requirement.
\end{itemize}
\end{itemize}
Following these rules, and supposing that our sample system were of assurance level 1, the 
test cases displayed in Example~\ref{ex:tcreqtrace} would be necessary. Test cases
$tc_{12}, tc_{13}, tc_{14}$, however, would not be required, since the TurnIndLvr has no impact on
{\sf REQ-002} according to the model: the risk of a hidden impact of this interface on the requirement
has already been taken into account when testing $tc_8, tc_9$.

\subsubsection{Test Strategies Proving Conformance}

An alternative for justifying test strategies consists in proving that they will finally 
converge to an exhaustive test suite establishing some conformance relation between model and SUT. 
This approach has a long tradition: one of the first contributions in this field was
Chow's W-Method~\cite{chow:wmethod} applicable for minimal state machines, which was generalised 
and extended into many directions, so that even in the core of the exhaustive test strategy 
for timed automata~\cite{Springintveld2001} some argument from the W-Method is used.

Though execution of   exhaustive
test suites will generally be infeasible in practice, convergence to exhaustive test suites ensures that
new test cases added to the suite will really increase the assurance  level by a positive amount: 
intuitively designed test strategies often do not possess this property, because additional test cases
may just re-test SUT aspects already covered by existing ones.   

The known exhaustive strategies typically operate on finite data types (discrete events, or
variables with data ranges that can easily be enumerated). It is an interesting research challenge
whether similar results can be obtained in presence of large data types, if application of 
equivalence class partitioning is justified. In~\cite{grieskamp2002} the authors formalise the concept
of equivalence class partitioning and prove that exhaustive suites can be constructed for white-box test
situations. In \cite{pel2012extended} this approach is currently generalised within the COMPASS project
with respect to black-box testing and  semantic models that are more general than the one underlying
the results presented in~\cite{grieskamp2002}.

%
%
%



\subsection{Challenges to Test Case Generation and Test Strategy Design}\label{sec:tcchallenge}


The size of SoS state spaces implies that exhaustive investigation of the complete 
concrete state space will certainly be infeasible. We suggest to
tackle this problem by two orthogonal strategies, as is currently investigated in the COMPASS project~\cite{D34.2}.
\begin{itemize}
\item On constituent system level, different behaviours associated with the same
 local mission threads\footnote{Mission threads are end-to-end tests; in the context described here, mission
 threads are executed on constituent system level.} will be comprised in equivalence classes. This reduces the complexity problem
 for SoS system testing to covering combinations of   classes of constituent system behaviours instead of sequences of concrete state vector combinations.
 
 \item On SoS system level, ``relevant'' class combinations are identified by means of different variants
 of impact analysis, such as data flow analyses or investigation of 
 contractual dependencies. Behaviours of constituent
 systems which do not affect the relevant class combinations under consideration will be selected 
 according to the principle of orthogonal arrays~\cite{Phadke&89}, because this promises an effective
 combinatorial distribution of unrelated behaviours exercised concurrently with the critical ones. 
\end{itemize}

Apart from size and complexity, SoS present another challenge, because they typically change their 
configuration dynamically during run-time.
The dynamic adaptation of test objectives is particularly relevant for run-time acceptance testing 
of changing SoS configurations. In contrast to development models for SoS, however, we only have to
consider bounded changes of SoS configurations, because every
test suite can only consider a bounded number of configurations anyway. It remains to investigate how
to determine configurations possessing sufficient error detection strength. Results from the field of 
mutation testing will help to determine this strength in a  systematic and measurable way.

A further problem for systems of SoS complexity is presented by the fact that not every behaviour
can be full captured in the model, which results in under-specification and non-determinism.
Test strategy elaboration in presence of this problem be achieved in the following way.
\begin{itemize}
\item The SoS system behaviour is structured into several top-level operational modes. It is expected that 
switching between these modes can be performed in a deterministic way for normal behaviour tests: it is
unlikely that SoS performing operational mode changes only on a random basis are acceptable 
and ``testworthy''.

\item Entry into failure modes is non-deterministic, but can be initiated in a deterministic way
for test purposes by means of pre-planned failure injections. 

\item The behaviour in each operational mode is not completely deterministic, but can be captured 
by sets of constraints governing the  acceptable computations in each mode. Test oracles will therefore 
no longer check for explicit output traces of the SUT but for compliance of the traces observed with
the constraints applicable in each mode. 

\item For test stimulation purposes the SMT solver computes sequences of feasible mode switches and
the test data provoking these switches.

\item Incremental test model elaboration can be performed by adding constraints identified during 
test observations to the modes where they are applicable. To this end, techniques from machine learning
seem to be promising.
\end{itemize}

Justification of test strategies will be performed by proving that they will ``converge'' to exhaustive
tests proving some compliance relation between SUT and reference model.

%

%
%
%
%
%
%
%


\section{Conclusion}\label{sec:conc}

In this article several aspects of industrial-strength model-based testing and its 
underlying methods have been presented. A reference tool has been described, so that the
presentation may serve as a benchmark for alternative tools capable of handling test campaigns
of equal or even higher complexity. Readers are invited to join the discussion on suitable benchmarks
for MBT tools -- initial suggestions on benchmarking can be found in~\cite{pel2011a} -- and to
contribute case studies and models to the MBT benchmark website {\tt http://www.mbt-benchmarks.org}.

A further topic beyond the scope of this paper is of considerable importance for tool builders: 
MBT tools automating test campaigns for safety-relevant systems have to be {\it qualified}, and
standards like RTCA DO-178C~\cite{DO178C} for the avionic domain, CENELEC EN650128~\cite{EN50128} for the railway domain, and ISO 26262~\cite{ISO26262-8} for the automotive domain have rather precise policies 
about how tool qualification can be obtained. 
A detailed comparison between tool qualification requirements of these standards is presented in~\cite{D34.1}, 
and it is described in~\cite{peleska2012c} how tool qualification has been obtained for RT-Tester.
We believe that the complexity of the algorithms required in MBT tools justifies that effort is spent on
their qualification, so that their automated application will not mask errors of the SUT due to undetected
failures in the tool. 

\paragraph{Acknowledgements.}
The author would like to thank the organisers of the MBT~2013 for giving him the opportunity to present the ideas summarised in this paper. Special thanks go to J{\"o}rg Brauer, Elena Gorbachuk, Wen-ling Huang, Florian Lapschies and Uwe Schulze for contributing  to the results presented here.
 
\bibliographystyle{eptcs}
\bibliography{jp,references}
\appendix
\section{Case Study: Turn Indication Control Function}\label{sec:turnind}

As a case study we consider the turn indication function of a vehicle providing left/right indication and emergency flashing by means of exterior lights flashing with a given frequency. Left/right indication is switched on by means of the turn indicator lever with its positions 0 (neutral), 1 (left), and 2(right). Emergency flashing is controlled by means of a switch with positions 0 (off) and 1 (on). 
Activating the indication lights is subject to the condition that the available voltage is sufficiently
high.
The requirements for the turn indication function are as shown in Table~\ref{tab:req}.

\begin{table}[htdp]
\footnotesize
\caption{Requirements of the turn indication control system}
\begin{center}
\begin{tabular}{|p{2in}|p{4in}|}\hline\hline
{\bf Requirement} & {\bf Description}
\\\hline\hline 
{\sf REQ-001} Flashing requires sufficient voltage &
Indication lights are only active, if the electrical voltage (input Voltage)
is $> 80\%$ of the nominal voltage.
\\\hline
{\sf REQ-002} Flashing with 340ms/320ms on-off periods &
If any lights are flashing, this is done synchronously with a 340ms ON -- 320ms OFF period.
\\\hline
{\sf REQ-003} Switch on turn indication left &
An input change from turn indication lever state TurnIndLvr = 0 or 2 to
TurnIndLvr = 1 switches indication lights left (output FlashLeft) into flashing mode and switches indication lights right (output FlashRight) off.
\\\hline
{\sf REQ-004} Switch on turn indication right &
An input change from turn indication lever state TurnIndLvr = 0 or 1 to
TurnIndLvr = 2 switches indication lights right (output FlashRight) into flashing mode and switches indication lights left (output FlashLeft) off.
\\\hline
{\sf REQ-005} Emergency flashing on overrides left/right flashing &
An input change from EmerFlash = 0 to EmerFlash = 1 switches indication lights left (output FlashLeft) and right (output FlashRight) into flashing mode, regardless of any previously activated turn indication.
\\\hline
{\sf REQ-006} Left-/right flashing overrides emergency flashing	 &
Activation of the turn indication left or right overrides emergency flashing, if the latter has been activated before.
\\\hline
{\sf REQ-007} Resume emergency flashing &
If turn indication left or right is switched off and emergency flashing is still active, emergency flashing is continued or resumed, respectively.
\\\hline
{\sf REQ-008} Resume turn indication flashing &
If emergency flashing is turned off and turn indication left or right is still active, the turn indication is continued or resumed, respectively.
\\\hline
{\sf REQ-009} Tip flashing &
If turn indication left or right is switched off before three flashing periods have elapsed, the turn indication will continue until three on-off periods have been performed.
\\\hline\hline
\end{tabular}
\end{center}
\label{tab:req}
\normalsize
\end{table}%

\begin{figure}
  \centering
 \includegraphics[width=6in]{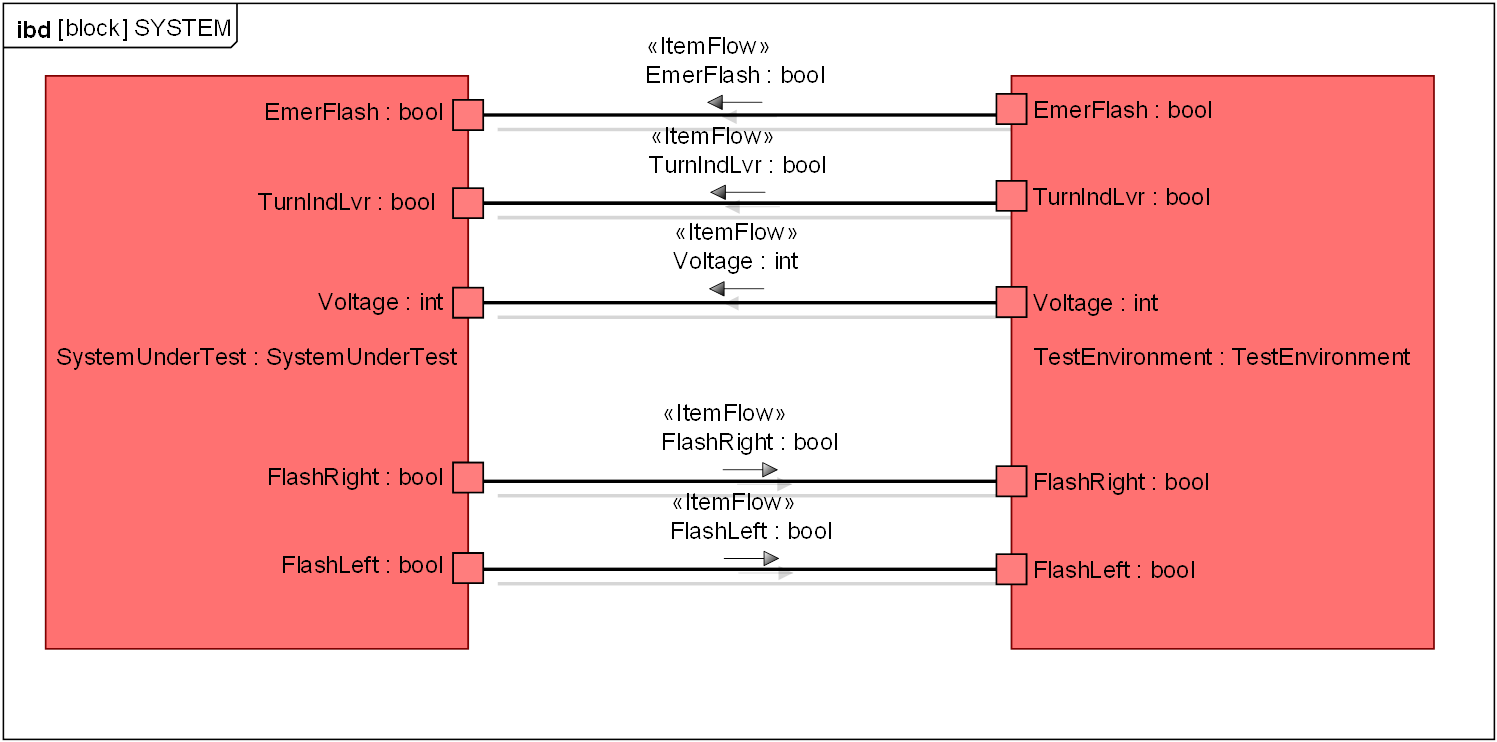}
  \caption{Interface between TE and SUT.}
  \label{fig:SYSTEM}
  \end{figure}

The SysML test model for this system structured  into TE and SUT blocks, as shown in Fig.~\ref{fig:SYSTEM}.
The interfaces shown in this diagram are the observable SUT outputs and writable inputs that may be accessed by the TE.
RT-Tester allows for SysML properties and signal events to be exchanged between SUT and TE model components. The tool provides interface modules mapping their valuations onto concrete software or hardware interfaces and vice versa.
In a software integration test  the turn indication lever values and the status of the emergency switch may be passed to the SUT, for example, by means of shared variables. The SUT outputs (left-hand side lamps on/off, right-hand side lamps on/off) can also be represented by Boolean output variables of the SUT. In a HW/SW integration test interface modules would map the turn indication lever status and the emergency flash button to discrete inputs to the SUT. In a system integration test the actual voltage and the current
placed by the SUT on the indication lamps would be measured. 
The interface abstraction required for the test level is specified by a signal map that associates abstract SysML model interfaces with concrete interfaces of the test equipment.

    \begin{figure}
  \centering
 \includegraphics[width=5in]{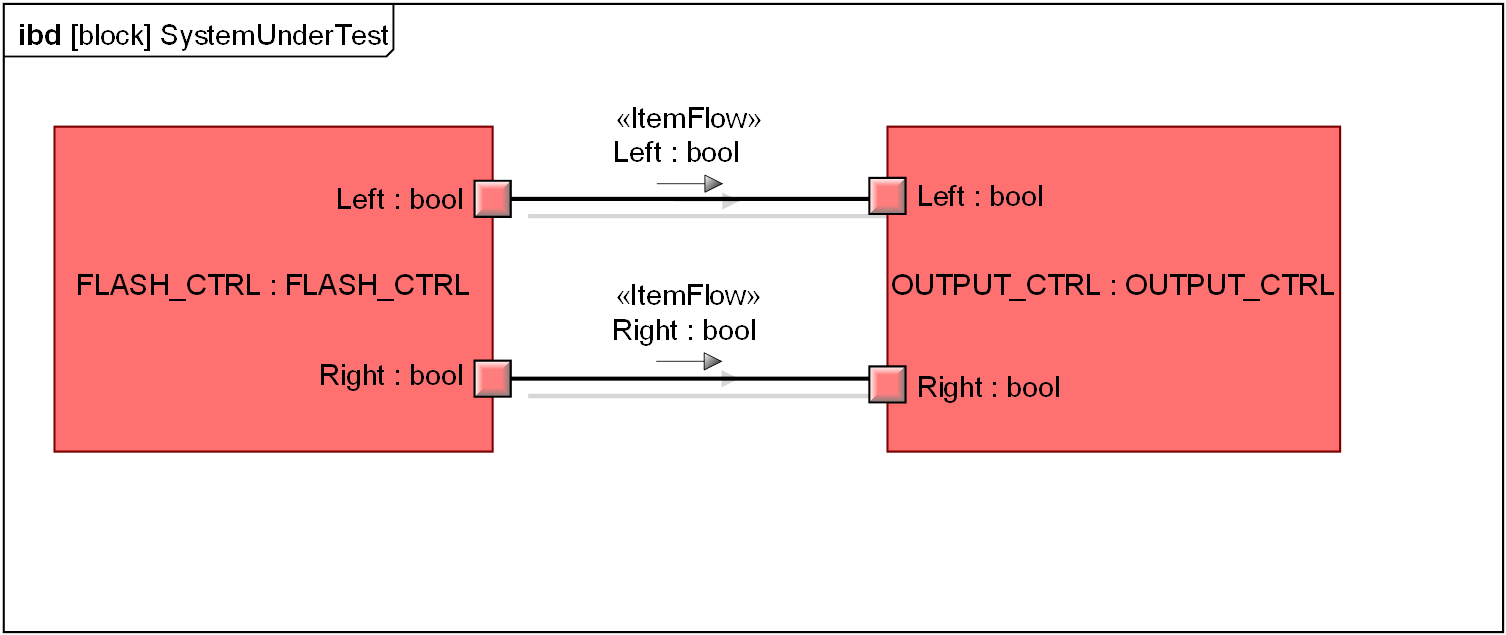}
  \caption{Functional decomposition of the SUT.}
  \label{fig:SUT}
  \end{figure}

\begin{figure}
  \centering
 \includegraphics[width=6in]{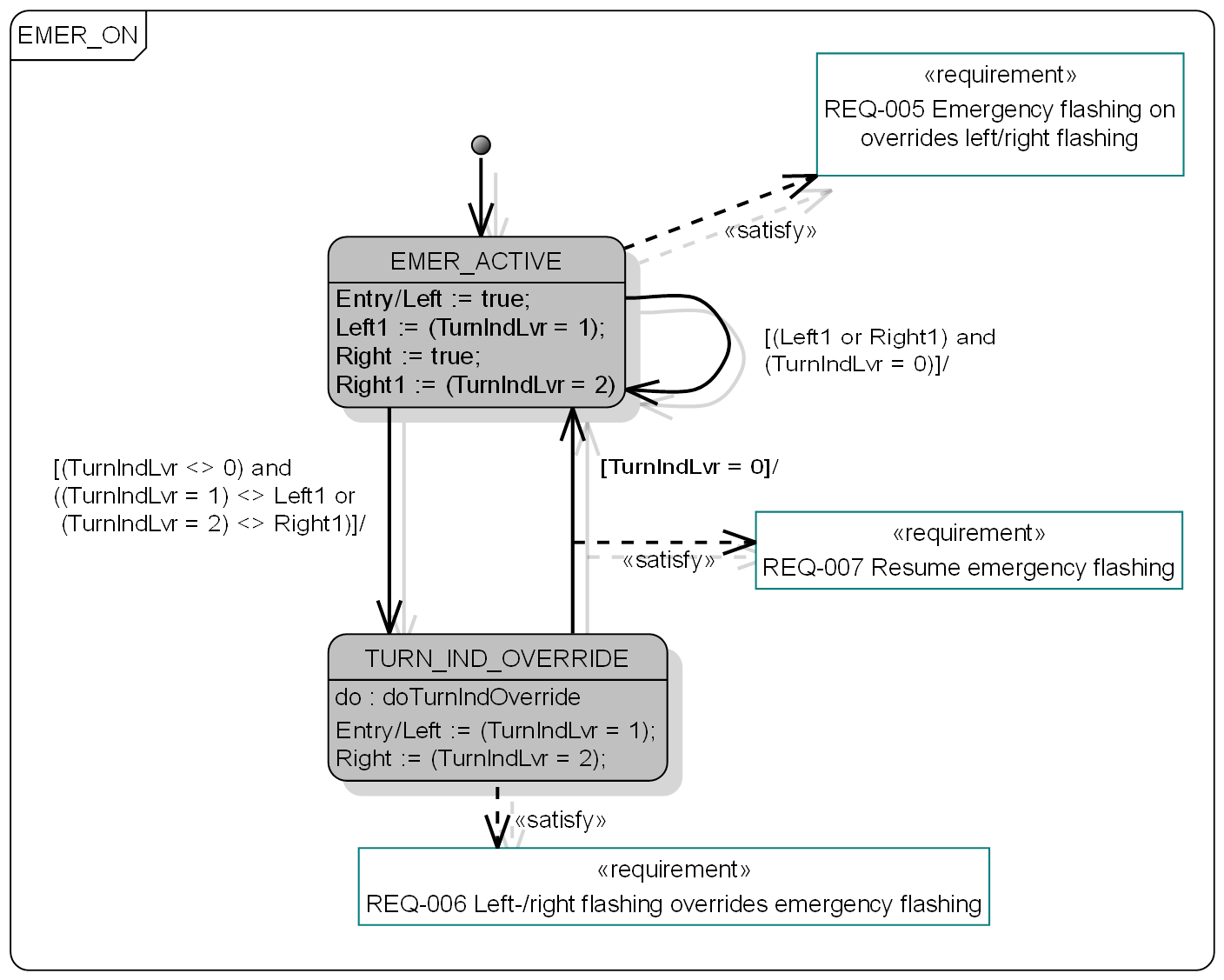}
  \caption{Decomposition of control state EMER\_ON.}
  \label{fig:EMER_ON}
  \end{figure}

The structural view on the SUT has to be decomposed further, until each block is associated with a sequential behaviour. For the case study discussed here, the SUT is further decomposed into two concurrent functions as depicted in Fig.~\ref{fig:SUT}.
 Functional component FLASH\_CTRL performs the decisions about left/right indication or emergency flashing.
 The decision is communicated to  component OUTPUT\_CTRL by means of internal interface Left (flashing on 
 left-hand side indication lights if Left = 1) and Right 
 (flashing on right-hand side indication lights if Right = 1). Block OUTPUT\_CTRL
  controls the flashing cycles and switches off indication lamps if the voltage gets too low.
 The FLASH\_CONTROL component operates as follows.
\begin{itemize}
\item As long as the emergency flash switch has not been activated, 
Left/Right are set according to the turn indication lever status. This is specified in do activity doEmerOff.
\item As soon as the emergency flash switch EmerFlash is switched on, Left/Right are set as specified in sub-state machine EMER\_ON (Fig~\ref{fig:EMER_ON}).
\item When entering EMER\_ON, 
Left/Right are both set to true and the state machine remains in control state EMER\_ACTIVE.
\item When the turn indication lever is changed to left or right position, 
emergency flashing is overridden, and left/right indication is performed.
\item Emergency flashing is resumed if the turn indication lever is switched into neutral position.
\end{itemize}

  \begin{figure}
  \centering
 \includegraphics[width=5in]{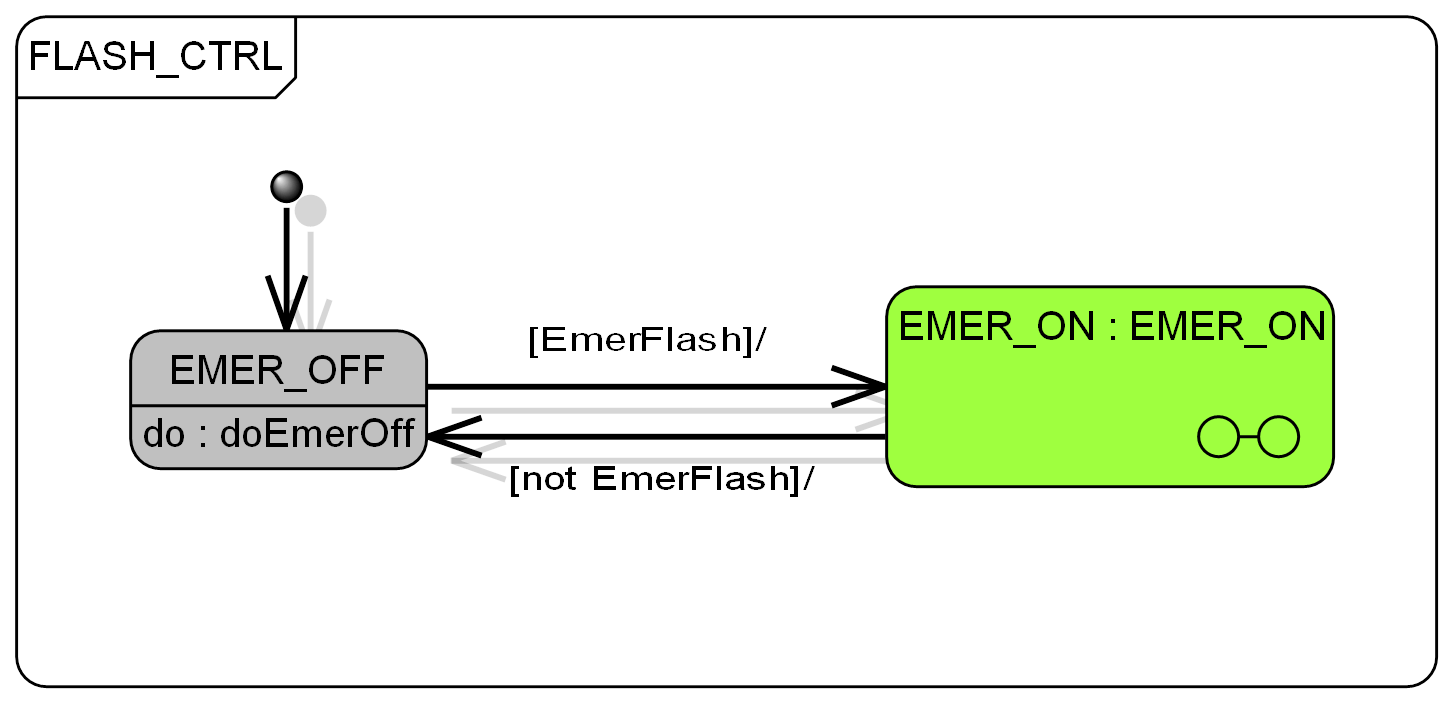}
  \caption{State machine controlling left/right and emergency flashing.}
  \label{fig:FLASH_CTRL}
  \end{figure}
  
\begin{figure}
  \centering
 \includegraphics[width=6in]{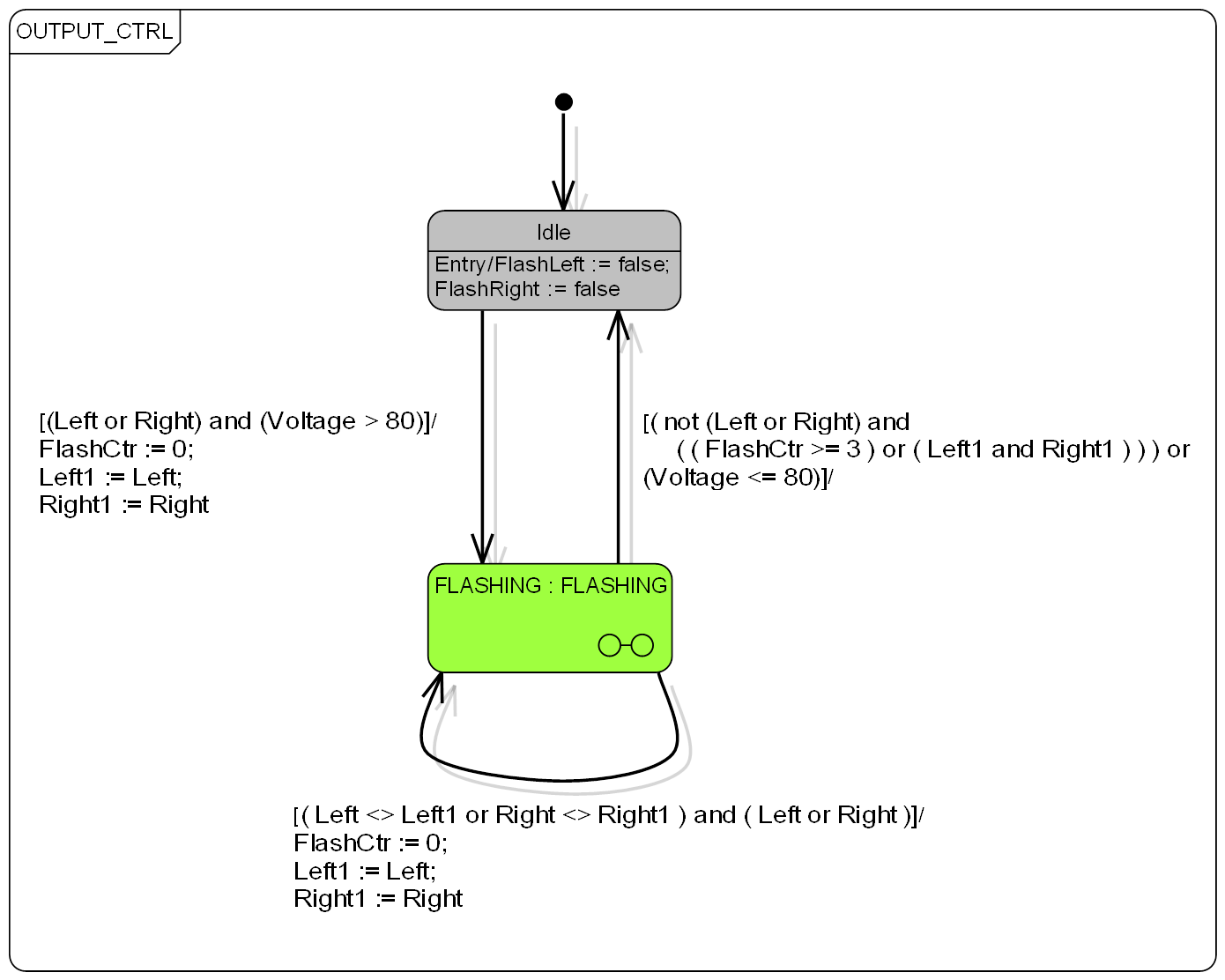}
  \caption{State machine switching indication lights.}
  \label{fig:OUTPUT_CTRL}
  \end{figure}

\begin{figure}
  \centering
 \includegraphics[width=6in]{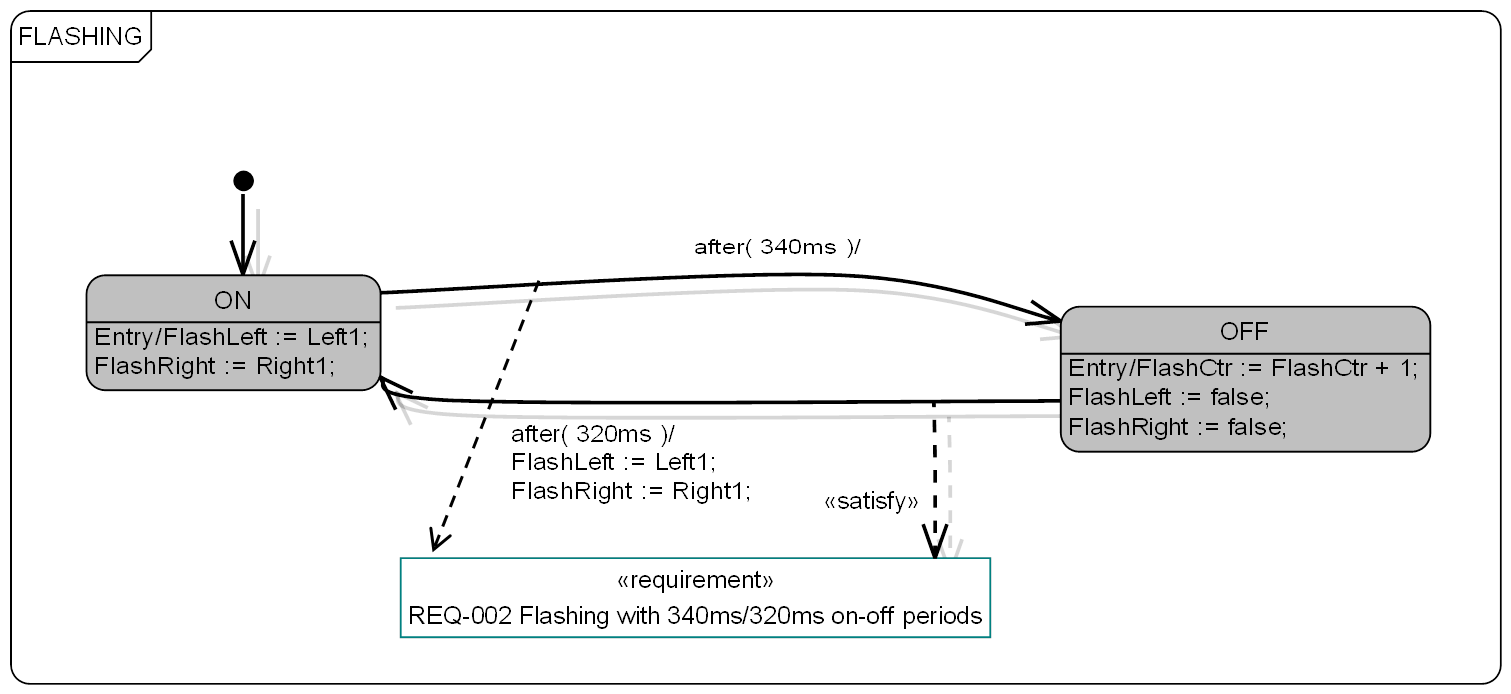}
  \caption{Decomposition of control state FLASHING.}
  \label{fig:FLASHING}
  \end{figure}

  Function OUTPUT\_CTRL sets the SUT output interfaces FlashLeft and FlashRight (Fig.~\ref{fig:OUTPUT_CTRL}
  and \ref{fig:FLASHING}). The indication lamps are switched according to the internal interface state Left/Right, if the voltage is greater than 80\% of the nominal voltage. After the lamps have been on for 340ms, they are switched off and stay so until 320ms have passed. A counter FlashCtr is maintained: if the turn indication lever is switched from left or right back to the neutral position before 3 flashing periods have been performed, left/right indication will remain active until the end of these 3 periods. 
  
  \begin{table}[htdp]
\caption{Requirements  and associated constraints identifying witness   computations.}
\begin{center}
\footnotesize
\begin{tabular}{|p{2.5in}|p{3.5in}|}\hline\hline
{\bf Requirement} & {\bf Constraint}
\\\hline\hline 
{\sf REQ-001} Flashing requires sufficient voltage &
{\sf \guillemotleft Constraint\guillemotright} 
$\tf (Voltage \leq 80)$
\\\hline
{\sf REQ-002} Flashing with 340ms/320ms on-off periods &
{\sf \guillemotleft Transition\guillemotright} $\text{ON} \fun \text{OFF}$  
{\sf \guillemotleft Transition\guillemotright} $\text{OFF} \fun \text{ON}$
\\\hline
{\sf REQ-003} Switch on turn indication left &
{\sf \guillemotleft Constraint\guillemotright} $\tf (\text{FlashLeft} = 1 \wedge \text{FlashRight} = 0)$
\\\hline
{\sf REQ-004} Switch on turn indication right &
{\sf \guillemotleft Constraint\guillemotright} $\tf (\text{FlashLeft} = 0 \wedge \text{FlashRight} = 1)$
\\\hline
{\sf REQ-005} Emergency flashing on overrides left/right flashing &
{\sf \guillemotleft Constraint\guillemotright} 
$\tf (\text{EMER\_OFF} \wedge \text{TurnIndLvr} > 0 \wedge \text{EmerFlash})$
\\\hline
{\sf REQ-006} Left-/right flashing overrides emergency flashing	 &
{\sf \guillemotleft Atomic State\guillemotright} TURN\_IND\_OVERRIDE
\\\hline
{\sf REQ-007} Resume emergency flashing &
{\sf \guillemotleft Transition\guillemotright}$\text{TURN\_IND\_OVERRIDE} \fun \text{EMER\_ACTIVE}$
\\\hline
{\sf REQ-008} Resume turn indication flashing &
{\sf \guillemotleft Constraint\guillemotright}\newline
 $\tf(\text{EMER\_ACTIVE} \wedge \neg \text{EmerFlash} \wedge \text{TurnIndLvr} > 0)$
\\\hline
{\sf REQ-009} Tip flashing &
{\sf \guillemotleft Constraint\guillemotright} \newline
$\tf(\text{Voltage} > 80 \wedge \neg (\text{Left}\vee \text{Right}) \wedge 
\text{Left1} + \text{Right1} = 1 \wedge \text{FlashCtr} < 3)$
\\\hline\hline
\end{tabular}
\normalsize
\end{center}
\label{tab:tracing}
\end{table}%

\end{document}